\documentclass[preprint,12pt]{elsarticle}
\usepackage{amsfonts,amscd,epsfig}
\usepackage{amssymb}
\journal{Physics Letters A}
\biboptions{square,comma}

\begin{document}
\begin{frontmatter}
\title{Scattering of quasi-one-dimensional solitons on
impurities in large Josephson junctions }
\author[is]{Ivan O. Starodub} 
\ead{starodub@bitp.kiev.ua}

\author[yz]{Yaroslav Zolotaryuk\corref{cor1}}
\ead{yzolo@bitp.kiev.ua}

\cortext[cor1]{Corresponding author}
\date{\today}
\address{
Bogolyubov Institute for Theoretical Physics,
National Academy of Sciences of Ukraine,
03680 Kyiv, Ukraine
}

\begin{abstract}
Fluxon transmission through impurities of different shape in a 
quasi-one-dimensional long Josephson junction is investigated. The 
junction width is significantly less than its length but, at the same 
time, is of the order of the Josephson penetration length $\lambda_J$
or exceeds it. The retrapping  
current on the impurities of the point, line and rectangular shape is 
computed as a function of the junction width both numerically and
 analytically.
Good agreement between the analytic formulae and the numerical 
simulation results for the intermediate (several $\lambda_J$) junction 
width is observed.
\end{abstract}
\end{frontmatter}

\section{Introduction}
\label{intro}

The dynamics of magnetic flux propagation in a long Josephson
junction (LJJ) has been and continues to be a subject of strong 
theoretical and practical interest during the last three
decades \cite{barone82,likharev86,u98pd}. The magnetic flux quantum
in a LJJ is a soliton (also known as {\it fluxon})
governed by the well-known sine-Gordon (SG) equation.
The convenient way to prepare a junction with the required 
properties is to install various inhomogeneities into it.
Up to now the substantial theoretical work has been devoted to the study of the
fluxon motion in the one-dimensional (1D) LJJs with point-like 
 \cite{ms78pra,kmn88jetp,mu90jap} and spatially
extended \cite{kkc88pla,km89jap,ddks11ejam} inhomogeneities. 
Experimental results on the fluxon scattering on impurities
are reported in \cite{mu90jap,expsize}
Spatially inhomogeneous Josephson systems with trapped fluxons 
have been discussed as prospective applications, such 
as fluxon-based information devices \cite{g05,fssk-s07prb}.

Real LJJs are always two-dimensional (2D), or, more precisely,
 quasi-one-dimensional (Q1D), in the sense that they have a finite
 width in the direction perpendicular to the direction of 
fluxon propagation. Up to now the fluxon dynamics in 
Q1D junctions has been
scarcely investigated as compared to the pure 1D case. 
Most of attention has been focused on the various isotropic 2D 
 structures like oscillons and ring  kinks \cite{co78pla-cl81pd}.
It is worth mentioning also the case of window 
junctions \cite{bcf02pd,cg04pc}, which can be called inverse in 
some sense: it studies point or rectangular junctions embedded in a larger
two-dimensional superconduncting sample.
Quasi-one-dimensional solitons, or, in other words, solitonic fronts
in infinite (in both $x$ and $y$ directions) samples
have been studied in detail by Malomed \cite{m91pd}.
Several interesting results in the absence of dissipation
have been reported including the waves in the Q1D sine-Gordon 
equation, travelling along the soliton line  
 \cite{gksyn} and skyrmion scattering
on impurities in the 2D baby Scyrme model  \cite{pzb05jpa}.
However, in the case of the fluxon dynamics in a LJJ the presence of 
dissipation is unavoidable.

It is of interest to investigate the Q1D fluxon dynamics in the presence
of spatial inhomogeneities when the junction width is finite. We expect
that the fluxon transmission in this case will be significantly enhanced
comparing to the pure 1D case. For example, in the previously
studied case of the lattice acoustic soliton front interaction with mass 
impurities \cite{zsc98prb} it has been shown that the front can
{\it round} the point impurity while a 1D
lattice soliton gets reflected from it. Moreover, the soliton front
can overcome even the impurity of the infinite mass.
To our knowledge the fluxon interaction with spatial inhomogeneities
has not been studied except in \cite{m91pd}, however this
paper deals only with the infinite sample width. 

Thus, in this Letter we aim at studying the Q1D fluxon interaction with
impurities and finding out how this interaction depends on the
junction width and other system parameters. In particular, it is planned
to find how the retrapping current (e.g., the minimal bias current for
which the fluxon propagation is still possible) depends
on the junction width.

The paper is organized as follows. In the next section, the model is 
described. Section \ref{sec3} 
is devoted to the fluxon transmission through
impurities. In the last section discussion and conclusions are presented.

\section{The model}
\label{model}

We consider the Q1D long Josephson junction (LJJ) subjected to the external
time-independent bias. The main dynamical variable is the difference
between the phases $\theta_2(x,y;t)-\theta_1(x,y;t)=\phi(x,y;t)$ of the
macroscopic wave functions of the superconducting layers of the
junction. The time evolution of the phase difference is governed by
the perturbed sine-Gordon (SG) equation of the form 
\begin{equation}\label{1}
\phi_{tt}-\Delta \phi +[1+f(x,y)]\sin \phi=-\alpha \phi_t-\gamma,
\end{equation}
where $\Delta \phi=\phi_{xx}+\phi_{yy}$ and the indexes ${t,x,y}$ stand
for the respective partial derivatives. In this dimensionless equation
the spatial variables $x$ and $y$ are normalised to
the Josephson penetration depth $\lambda_J$, the temporal variable $t$
is normalised to the inverse
Josephson plasma frequency $\omega_J^{-1}$ \cite{barone82,likharev86}.
The bias current $\gamma$ is normalised to the critical Josephson
current of the junction and $\alpha$ is the dimensionless 
 dissipation parameter.
The function $f(x,y)$ describes the spatial inhomogenity. In the case
of point impurities in the general form it reads
\begin{equation} \label{2}
f(x,y)=\sum_{n=1}^N \mu \delta (x-a_n)\delta (y-b_n)\:.
\end{equation}
It is supposed that there are $N$ impurities
in this junction,
positioned at the points $x=a_n$, $y=b_n$, $n=1,2,\ldots,N$,
 with $\mu$ being the ``strength'' or the amplitude
of the impurity. Only the microshorts ($\mu>0$), i.e., a narrow 
regions of locally enhanced critical
density of the tunnelling superconctucting current will be investigated
in this article.
However, the size an inhomogeneity
in experimental samples is finite \cite{expsize}. Therefore, we consider
also the case of the line microshort of width $d_y$, stretched
along the $y$ direction:
\begin{equation}\label{3}
f(x,y)=\mu \delta(x) \left [\theta\left (y+\frac{d_y}{2}\right )+
\theta \left (\frac{d_y}{2}-y\right ) \right ] \,.
\end{equation}
Here $\theta(x)$ is the Heavyside function.
And finally, the rectangular impurity of the finite size in both $x$ and
$y$ directions 
\begin{equation}\label{4}
f(x,y)={\mu} \left [\theta\left (x+\frac{d_x}{2} \right)+
\theta\left (\frac{d_x}{2}-x\right)\right ]
\left [\theta\left (y+\frac{d_y}{2}\right )+
\theta \left (\frac{d_y}{2}-y\right ) \right ]
\end{equation}
will be considered as well.
It should be noted that the impurity strength $\mu$ has different
meanings in all three cases (\ref{2})-(\ref{4}). For the point impurity (\ref{2}) 
setting $w=0$
does not automatically yield the pure 1D case studied before in
\cite{ms78pra,kmn88jetp,mu90jap}. This case can be retained in the
strip impurity case if $d_y=w$. The 1D finite-size impurity
case \cite{kkc88pla,km89jap,ddks11ejam} is retained in the same way.

We choose the boundary conditions along the $y$ direction 
in the von Neumann form:
\begin{equation}\label{5}
\phi_y \left(x,-{w \over 2},t\right)=\phi_y\left (x,{w \over 2},t
\right )=0.
\end{equation}
The boundary conditions along the $x$ axis are periodic: 
$\phi(x+L,y;t)=\phi(x,y;t)+2\pi$, where $L$ is the junction length, 
$L>w \gg 1$.

\section{Quasi-one-dimensional fluxon transmission through an impurity}
\label{sec3}

\subsection{Numerical simulations}

\begin{figure}[htb]
\centerline{\epsfig{file=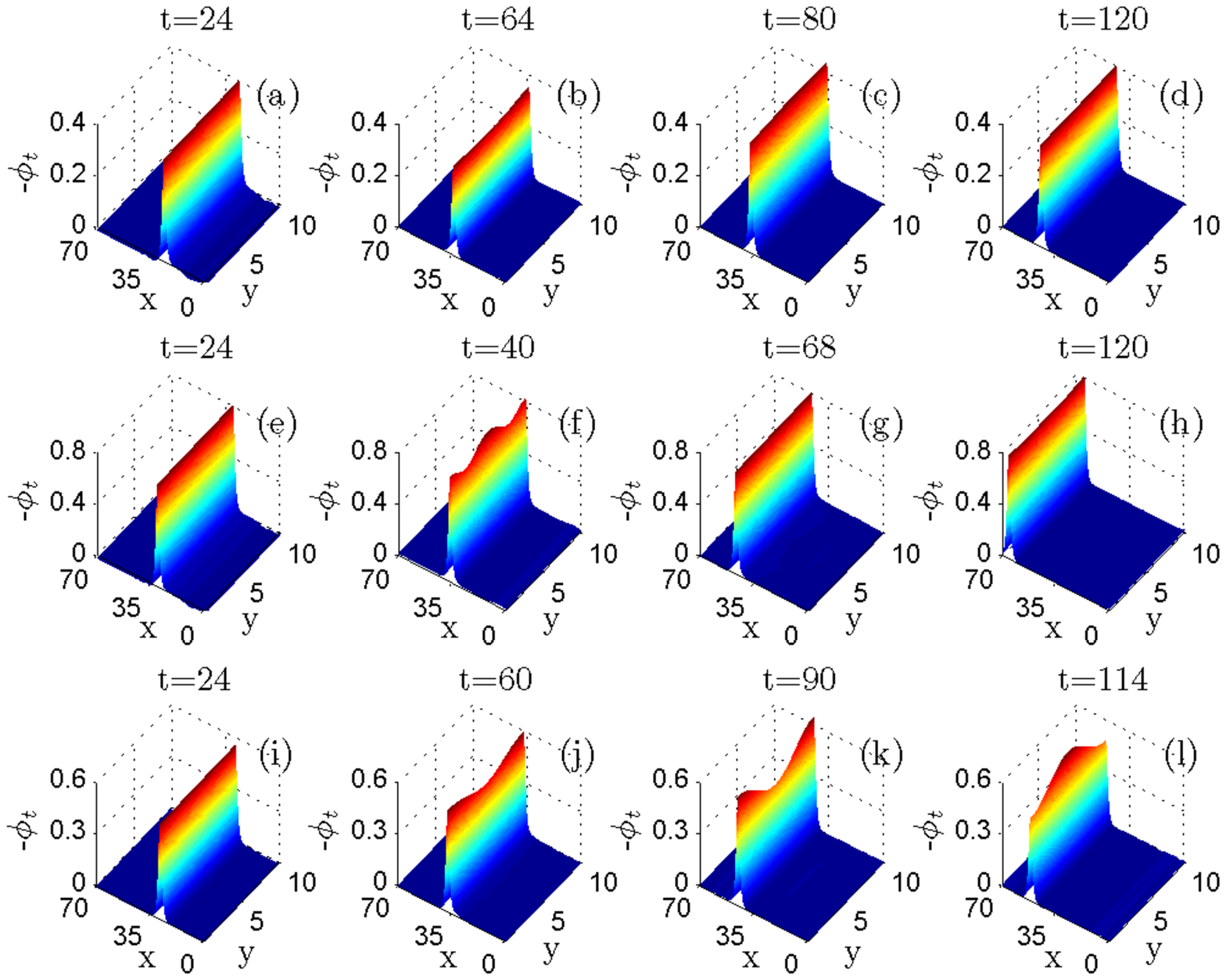,width=7.3in,angle=0}}
\caption{(Colour online) Time evolution of the Josephson phase derivative 
in the $L=70 \times w=10$ junction with 
$N=2$ impurities with the distance $a_y=4$ between them at $\mu=0.5$ and $\alpha=0.1$, $\gamma=0.03$ [(a)-(d)],
$\alpha=0.05$, $\gamma=0.02$ [(e)-(h)], $\alpha=0.01$, $\gamma=0.0035$ [(i)-(l)].
}
\label{fig1}
\end{figure}

In order to get an idea about the character of the fluxon
dynamics, the numerical integration of the Q1D 
 SG equation (\ref{1}) has been performed. The Josephson 
phase and its space derivatives are
discretized in the following way: $\phi(x,y;t) \to \phi(mh,nh;t)\equiv \phi_{mn}(t)$,\\
$\Delta \phi \simeq h^{-2}\left (\phi_{m+1,n}+\phi_{m,n+1}+
\phi_{m-1,n}+\phi_{m,n-1}-4 \phi_{mn}\right )$, while 
the $\delta$-function is approximated by the Kronecker $\delta$-symbol.
The resulting set of ODE's with the boundary conditions were
integrated using the 4th order Runge-Kutta scheme.
Details of the fluxon propagation through the two identical point
impurities (\ref{2}) placed along the $y$ axis at $a_1=a_2=0$ and 
$b_1=-b_2=2$ are given in Fig. \ref{fig1}.
It is important to mention that the dissipation in Eq.~(\ref{1}) 
is crucial and the soliton interaction
with impurities differs from the dissipationless case where the 
complex resonant behaviour occurs either for the $\delta$-like \cite{ksckv92jpa} or for the finite-size \cite{a-az08jpa} obstacles.
Far away from the impurities the fluxon exists as only one attractor
of the system with the velocity, predefined by the
damping parameter and external bias. 
Therefore, contrary to the non-dissipative case, 
the transmission consists of only 
two possible scenarios: trapping and passage.

For the sake of better visualization the derivative $-\phi_t$ is
plotted on the $xy$ plane for the different time moments and for
three different dissipation values: $\alpha=0.1$, $\alpha=0.05$ and 
$\alpha=0.01$. Without loss of generality the topological charge 
is assumed to be $Q=1$ (soliton) throughout the paper. 
The initial conditions are taken in the form of 
the approximate (invariant in the $y$ direction) soliton 
solution, placed at the beginning of the junction and having with the equilibrium velocity. 
%
%

\begin{figure}[htb]
\centerline{\epsfig{file=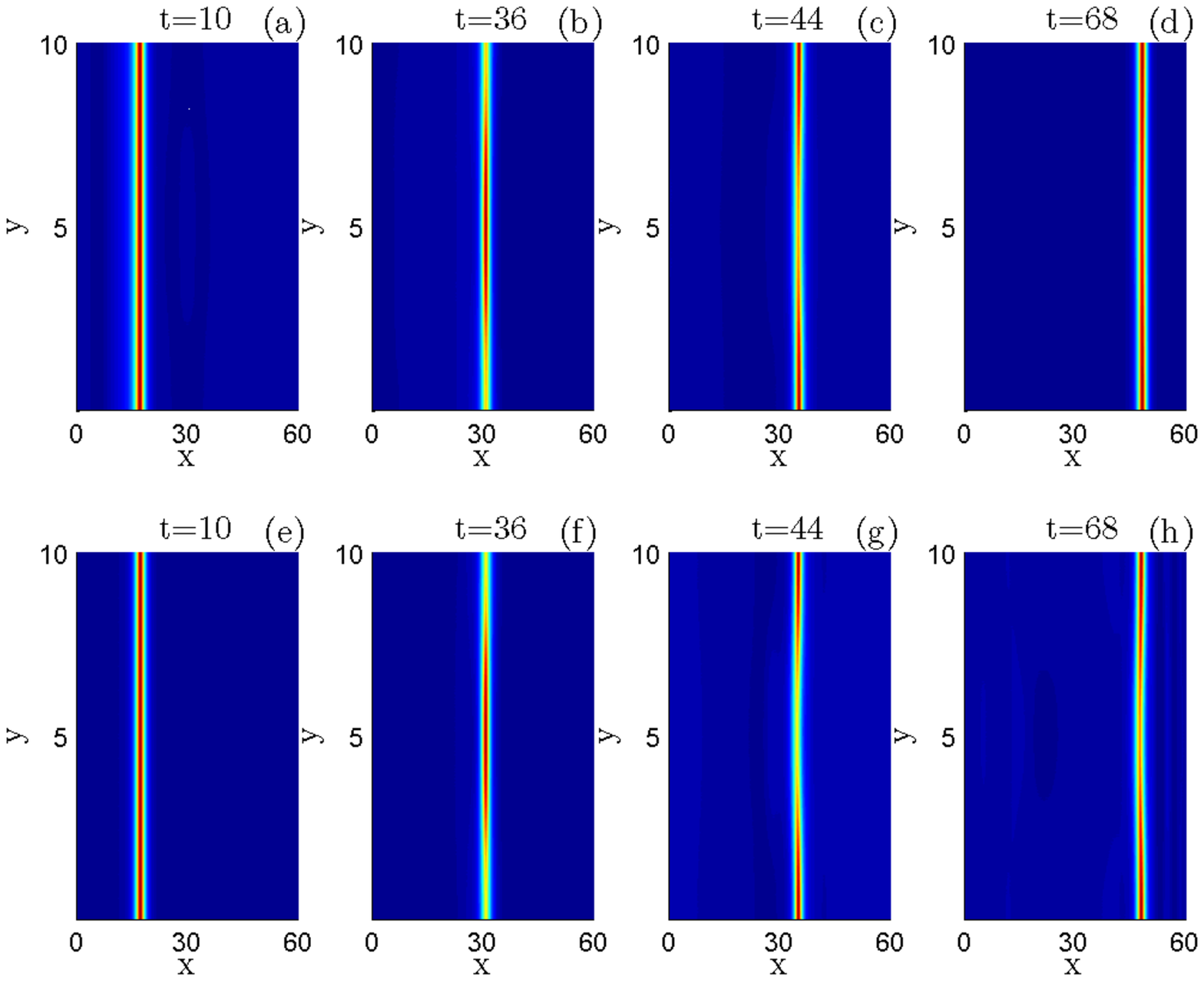,width=7.3in,angle=0}}
\caption{(Colour online) Contour plot of the time evolution of 
the Josephson phase derivative 
in the $L=60 \times w=10$ junction with the rectangular impurity 
(\ref{4}) with $\mu=0.5$, $d_x=3$, $d_y=5$, $\alpha=0.1$, $\gamma=0.08$ [(a)-(d)]
and $\alpha=0.01$, $\gamma=0.008$ [(e)-(h)].
}
\label{fig2}
\end{figure}

The fluxon interaction with the rectangular impurity is presented in
Fig. \ref{fig2} for $\alpha=0.1$ and $\alpha=0.01$. In this case,
the countour plot of the function $-\phi_t$ is plotted on the
$xy$ plane.
%
The fluxon retrapping current was computed numerically as well,
showing its steady decay when the junction width was increased.
These results will be discussed in detail in the next subsection.

The following conclusions can be drawn from these results. The
fluxon shape experiences certain changes after passing the impurities, namely the redistribution of the Josephson phase along the fluxon
line in the $y$ direction, as well as the
slight bending of the fluxon shape in the same direction.
These distortions eventually die out after some time, for 
$\alpha=0.05$ and $\alpha=0.1$ this happens quite soon after the passage through the impurity. For smaller dissipation ($\alpha=0.01$, as shown 
in Figs. \ref{fig1}i-l) the oscillations of the Josephson
 phase along the $y$ direction seem to survive for much longer time,
 comparable with the fluxon propagation time along the junction. 
The numerical simulations for thinner junctions,
$w<10$ (not shown in the paper), demonstrate that these shape
distortions are much weaker and are hardly noticeable. 
Thus, when studying the Q1D fluxon interaction with impurities
one can assume with the high degree of certainty that the fluxon is
 an almost hard rod at least if $w \sim {\cal O}(1)$. Due to finiteness
of the junction in the $y$ direction and the boundary conditions (\ref{5}),
the straight soliton front is the energetically most favourable
solution and thus it is not possible to observe 
the arc-like solitons reported in \cite{m91pd}.

\subsection{Retrapping current calculation}

Similarly to fluxon propagation in the 1D LJJ \cite{ms78pra,kmn88jetp,mu90jap}, there must exist two characteristic 
values of the bias current, $\gamma_c$ and 
$\gamma_{thr}$, $\gamma_c > \gamma_{thr}$. Moreover, the 
current-voltage
 characteristics of the LJJ with impurities have hysteretic 
 nature \cite{kmn88jetp,mu90jap}.
If $\gamma > \gamma_c$, the pinning on the impurity is not possible and there exists only one attractor that 
corresponds to fluxon propagation. This happens because
the bias current is too strong for the fluxon to get trapped on the
impurity.  In the interval 
$\gamma_{thr}< \gamma < \gamma_c$, at least two attractors coexist: one corresponds
to fluxon pinning on the impurity (there can be several
different pinned fluxons if the impurity has finite length \cite{ddks11ejam}) and another one to fluxon 
propagation. If $\gamma < \gamma_{thr}$, the only possible regime is 
fluxon pinning on the impurity. 
The value of $\gamma_c$ is
defined only by  the properties of the impurity,
and can be obtained directly from the 1D analog. 
Contrary, for the retrapping current the dimensionality of the
junction and its width are crucial. 

Far from the impurity, the fluxon kinetic energy is proportional 
to the junction width and equals
$E_k=8w[(1-v_\infty^2)^{-1/2}-1]$. Here, $v_{\infty}=[1+(4\alpha/(\pi\gamma ))^2]^{-1/2}$ is the equilibrium fluxon velocity in the spatially
 homogeneous LJJ \cite{ms78pra}. In the non-relativistic fluxon case ($|v_\infty|\ll 1$), 
one gets $v_{\infty}\simeq \pi\gamma/(4\alpha)$.
By substituting the ansatz \\
$\phi(x,y;t)=4\arctan \;\exp \{\left [x-X(y,t) \right ]/\sqrt{1-v^2}\}$
into Eq. (\ref{1}), where  $X(y,t)$ is the coordinate
of the fluxon center of mass, one can obtain the 
Newtonian equation of motion for the fluxon center. 

Since the fluxons under consideration are extended objects in the
$y$ direction, the equation for the center of mass dynamics
as well as the impurity potential should depend on $y$. 
Taking into account the numerical simulation in the previous
section, we consider the fuxon as an absolutely rigid rod. 
We also mention that the impurity function $f(x,y)$ can be factorised 
in the cases (\ref{3})-(\ref{5}), therefore its
center of mass dynamics can be effectively projected on the $x$
axis and the respective equation of motion can be written as
\begin{equation}\label{7}
m {\ddot X} + m\alpha {\ddot X} + \frac{\partial U(X)}{\partial X}=0,
~U(X)=-2\pi \gamma X+U_0(X)\;, 
\end{equation}
where the center of mass coordinate $X$ depends only on time.
The impurity potential $U_0(X)$ now can be calculated from the
respective 1D problem by simply taking away the $y$-dependent
part in Eqs. (\ref{2})-(\ref{4}).
The fluxon mass $m$ has to be rescaled depending on the type
of the impurity. In the pure 1D case $m=8$. 
This assumption works well only if the impurity is consists
of lines and/or rectangles. If it
has a more complex shape, for example, like triangle, the
projection on the 1D problem becomes more complicated.

{\bf Point impurity.}
In the point impurity case (\ref{2}), we consider the line of 
$N$ identical equidistant impurities with $a_n=0$, $n=1,2,\ldots,N$
which are placed symmetrically with respect to the central line
$y=0$. Each impurity creates the potential
$2\mu /\cosh^2{X}$. In order to project the problem on the 1D picture 
it is necessary to rescale the fluxon mass from $m=8$ 
to $m=8w/N$, thus within the kinematic approach 
the retrapping current can be found as
a root of the energy balance equation $E_k=2\mu$, where $E_k=m
[(1-v_\infty^2)^{-1/2}-1]\simeq 4wv_\infty^2/N+{\cal O}(v_\infty^4)$ is 
the fluxon kinetic energy. In the non-relativistic case one
gets $\gamma_{thr}\simeq ({\alpha}/{\pi})\sqrt{{8\mu N}/{w}}$.
The correction of the order ${\cal O}(\alpha^2)$ can be taken into account 
with the help of the method, developed in \cite{kmn88jetp}. 
The modification of this method for the Q1D case is straightforward, 
therefore we describe only the main
steps.  The improved energy balance relation equates the
fluxon energy at $X=-\infty$ and its losses due to the dissipation, 
$\Delta E$, with the maximal height of the potential barrier $U(X)$:
\begin{equation}\label{en-bal}
m \left ({\pi\gamma\over {2\alpha}} \right)^2+\Delta E=U(X_{max}),
\end{equation}
where $X_{min}$, $X_{max}>X_{min}$ are the extrema of the potential
$U(X)$, $X_{min} \simeq -\ln(\mu/\gamma)/2$, 
$X_{max} \simeq - \pi\gamma/(2\mu)$.
 Equation (\ref{en-bal}) is universal and will be used for
all types of impurities.
The energy loss due to the dissipation equals \\
$\Delta E=8\alpha wN^{-1} \int_{-\infty}^{X_{max}}(v_\infty-\dot X)dX \simeq
4 \ln 2\sqrt{2\mu w} \alpha $.  Inserting this correction 
term $\Delta E$
into the improved energy balance equation and keeping the terms
up to the order ${\cal O}(\alpha^2)$, one gets the
 final corrected expression for the retrapping current: 
\begin{equation}\label{4a}
\gamma_{thr}\simeq \frac{\alpha}{\pi}\left (
\sqrt{\frac{8\mu N}{w}}-4\alpha \ln 2 \right ).
\end{equation}
 It appears that the $\gamma_{thr}(w)$
 relation is obtained simply by dividing the impurity
 strength by the factor $w/N$, moreover, the ${\cal O}(\alpha^2)$ correction
 does not depend on the junction width at all. 
From this expression one can clearly see that if only 
the ${\cal O}(\alpha)$ 
term is taken into account, the retrapping current
disappears as $w\to \infty$, thus, in the infinitely wide junction a
 fluxon always passes the impurity. The second term in Eq.~(\ref{4a})
does not depend on $w$, and, therefore it may lead to the wrong conclusion
that $\gamma_{thr}$ does not tend to zero as $w\to \infty$. However,
it should be noted that this term has been derived under the assumption
of $w$ being finite. 

The numerical simulations confirm the main suggestion of this Letter:
the retrapping current decays with the growth of the junction
width. The approximate expression (\ref{4a}) appears to be in good
agreement with the numerical data for smaller values of $w$, while
for larger $w$\rq{s} the numerical and analytic data diverge
(however, the agreement remains to be satisfactory). All this
is presented in Fig. \ref{fig1}a where the analytic results are given
by solid lines and the numerical data are shown by markers.
In the case of two and three impurities, placed along the line in
the $y$ direction, the retrapping
current virtually does not depend on the distance $a_y$ between them. For 
instance, the markers that correspond to $a_y=2$ ($\times$),
$a_y=3$ ($\circ$), $a_y=4$ ($\blacklozenge$) at $N=2$ and $w=5$ 
are almost indistinguishable. 

One may consider the excentrically placed (i.e., lying away from the
$x=0$ axis) impurity. 
The kinematic approach does not distinguish the 
impurities, placed at the different positions. The numerical simulations
show that the difference between the retrapping currents in both
the cases is very small. For example, $\gamma_{thr}=0.03770$
for the centrally placed impurity, $a_1=b_1=0$, 
at $w=2$ (the rest of the parameters are
 as in Fig. \ref{fig1}a) and $\gamma_{thr}=0.03776$ for the impurity
at $a_1=0$, $b_1=w/4$~.
%
%
\begin{figure}[htb]
\centerline{\epsfig{file=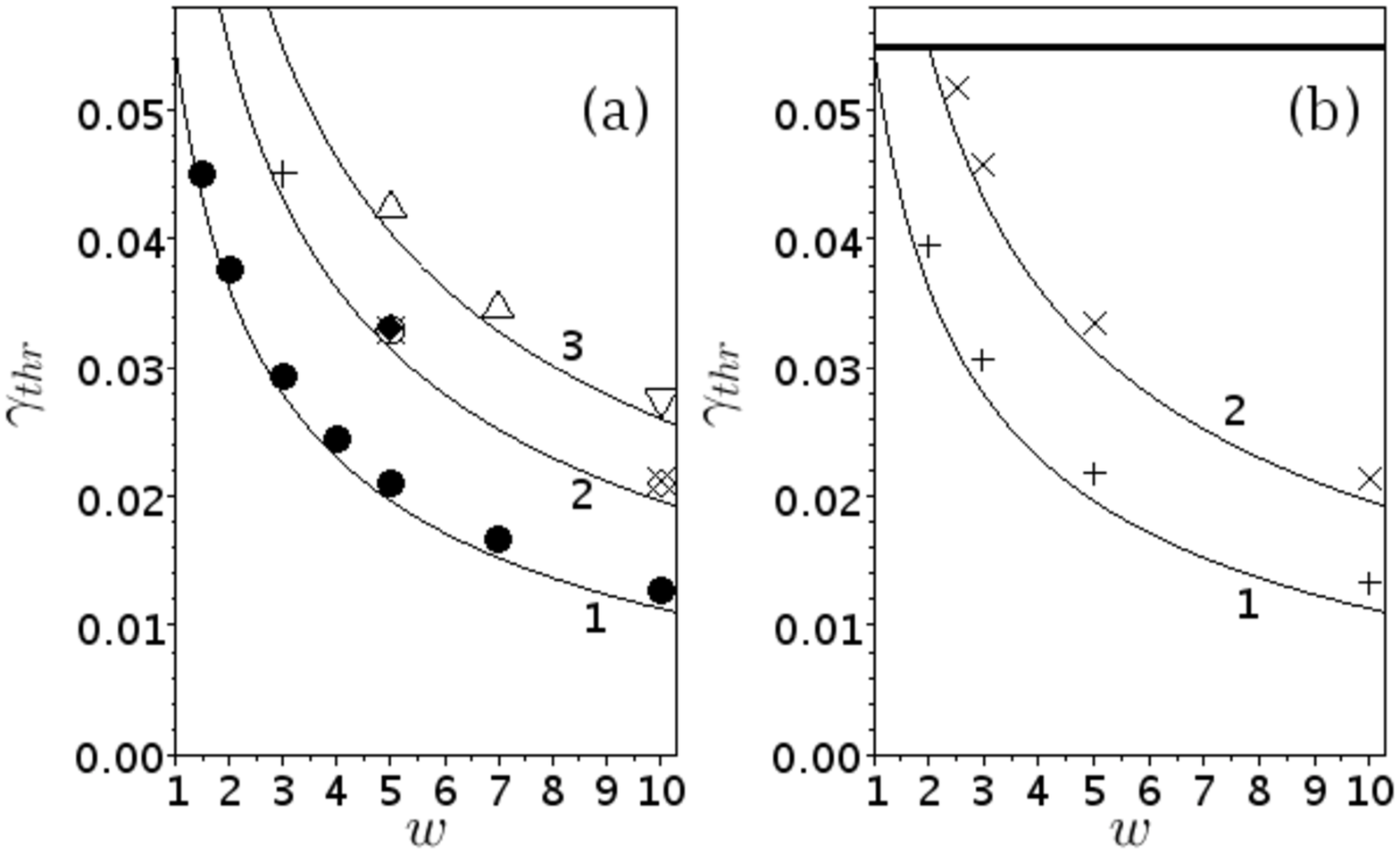,width=5.9in,angle=0}}
\caption{Retrapping current as a function of the junction width $w$
for $\alpha=0.1$, $\mu=0.5$.
Panel (a) corresponds to the case of point impurities (\ref{2}) with
solid lines illustrating the
approximate formula (\ref{4a}) at $N=1$ (line 1), $N=2$ (line 2) and
$N=3$ (line 3).
Markers correspond to the results of numerical simulations at
$N=1$ ($\bullet$); $N=2$ with $a_y=1$ ($+$), $a_y=2$ ($\times$),
$a_y=3$ ($\circ$), $a_y=4$ ($\blacklozenge$) and 
$a_y=7$ ($\diamond$); $N=3$ with $a_y=1$ ($\Delta$) and $a_y=2$ ($\nabla$).
Panel (b) corresponds to the case of the line impurity (\ref{3})
with $d_y=1$ (1 and $+$) and $d_y=2$ (2 and $\times$). Markers correspond
to numerical results and solid lines correspond to the formula
(\ref{4b}). Thick horizontal line corresponds to the retrapping current
on the point microshort in the pure 1D case.
}
\label{fig3}
\end{figure}

{\bf Line impurity. } In the case of the impurity line (\ref{3}), 
the same energy balance expression holds as for the point
impurity but the fluxon mass is assumed to be $m=8w/d_y$.
The rest of the calculation procedure for the retrapping current
is the same as for the point impuritiy. As a result, we obtain the 
the final expression for the retrapping current:
\begin{equation}\label{4b}
\gamma_{thr}=\frac{\alpha}{\pi} \left (
\sqrt{\frac{8\mu d_y}{w}}-4\alpha \ln 2 \right )~.
\end{equation}
Numerically computed $\gamma_{thr}$ appears to be in good correspondence
with the approximate expression (\ref{4b}) as shown in Fig. \ref{fig3}b. 
Similarly to the
point impurity case, the discrepancy between the analytical and
numerical results increases at larger $w$. In the limit $d_y/w \to 1$,
the effective 1D picture is restored because the impurity strip crosses
the whole junction in the $y$ direction. Thus, the retrapping current 
attains the value $\gamma_{thr}={\alpha}(
\sqrt{8\mu}-4\alpha \ln 2 )/\pi$ for the 1D soliton case (shown by the
thick horizontal line in Fig. \ref{fig3}b).  Thus, we observe that
if the strip impurity consitutes, for example, about $1/3$ of the 
junction width, the retrapping current is about $40\%$ less than the
respective 1D value.

{\bf Rectangular impurity.}
Finally, we consider the rectangular impurity (\ref{4}). In the point-particle description, the fluxon ``feels'' this impurity 
as the potential
$U_0(X)=2\mu [\tanh(X+d_x/2)+\tanh(d_x/2-X)]$ \cite{kkc88pla}. In this
case, the energy balance reads
$8w[(1-v_\infty^2)^{-1/2}-1]=4\mu d_y \tanh (d_x/2)$, from which
the expression for the threshold current can be easily computed.
For large $d_x$ it can be complemented by the correction 
 ${\cal O}(\alpha^2)$ 
computed for the 1D case in \cite{kkc88pla,km89jap}. 
The fluxon mass should be renormalized as $8 \to 8 w/d_y$.
In this limit it is assumed that
the impurity creates the step-like potential $U_0(X)=2 \mu [\tanh X+1]$,
which has a minimum at $X_{min}=-\mbox{arcsech}\sqrt{\pi\gamma/\mu}$ 
and a maximum at $X_{max}=\mbox{arcsech}\sqrt{\pi\gamma/\mu}$.
The energy losses due to the dissipation
$\Delta E=8\alpha w d_y^{-1} \int_{-\infty}^{X_{max}}(v_\infty-{\dot X})dX \simeq
4\sqrt{\mu d_y w^{-1}}\alpha \{\ln [\mu/(\pi\gamma)]+{\cal O}(1)\}$ should be substituted in the energy balance equation (\ref{en-bal}).
The final approximation reads
\begin{equation}\label{9}
\gamma_{thr}=\frac{\alpha}{\pi}\sqrt{\frac{d_y}{w}}\left [
4\sqrt{\mu \tanh (d_x/2)}-\alpha \ln \left (
{\mu \over \alpha^2}\frac{w}{d_y} \right ) \right ].
\end{equation}
Note that at variance with the point and line impurities, the
second correction depends on the both junction and the impurity width.
In the 1D limit ($d_y/w \to 1$), the formula (\ref{9}) retains
the form, found in \cite{kkc88pla,km89jap}.
In the limit $w \to \infty$, the second term decays to zero, but
much slower than the decay law ${\cal O}(w^{-1/2})$ of the first term.
Once again, the second term is a correct approximation only
for finite values of $w$.

In the opposite limit, $d_x\to 0$, one can use the following
approximation: $\tanh(X+d_x/2)+\tanh(d_x-X)=\sinh d_x/[\cosh^2 X+
\sinh^2 (d_x/2)]\simeq d_x \mbox{sech}^2 X +{\cal O}(d_x^2)$. Thus, 
the effective potential $U(X)$ is approximately the same as for the 
line impurity up to the coefficient $\sinh {d_x} \simeq d_x$.
Therefore the formula (\ref{4b}) can be modified accordingly and
finally the retrapping current reads
\begin{equation}\label{10}
\gamma_{thr}=\frac{\alpha}{\pi}\left [
4\sqrt{\frac{\mu d_y d_x/2}{w}}-4\alpha\ln 2 \right ].
\end{equation}

\begin{figure}[htb]
\centerline{\epsfig{file=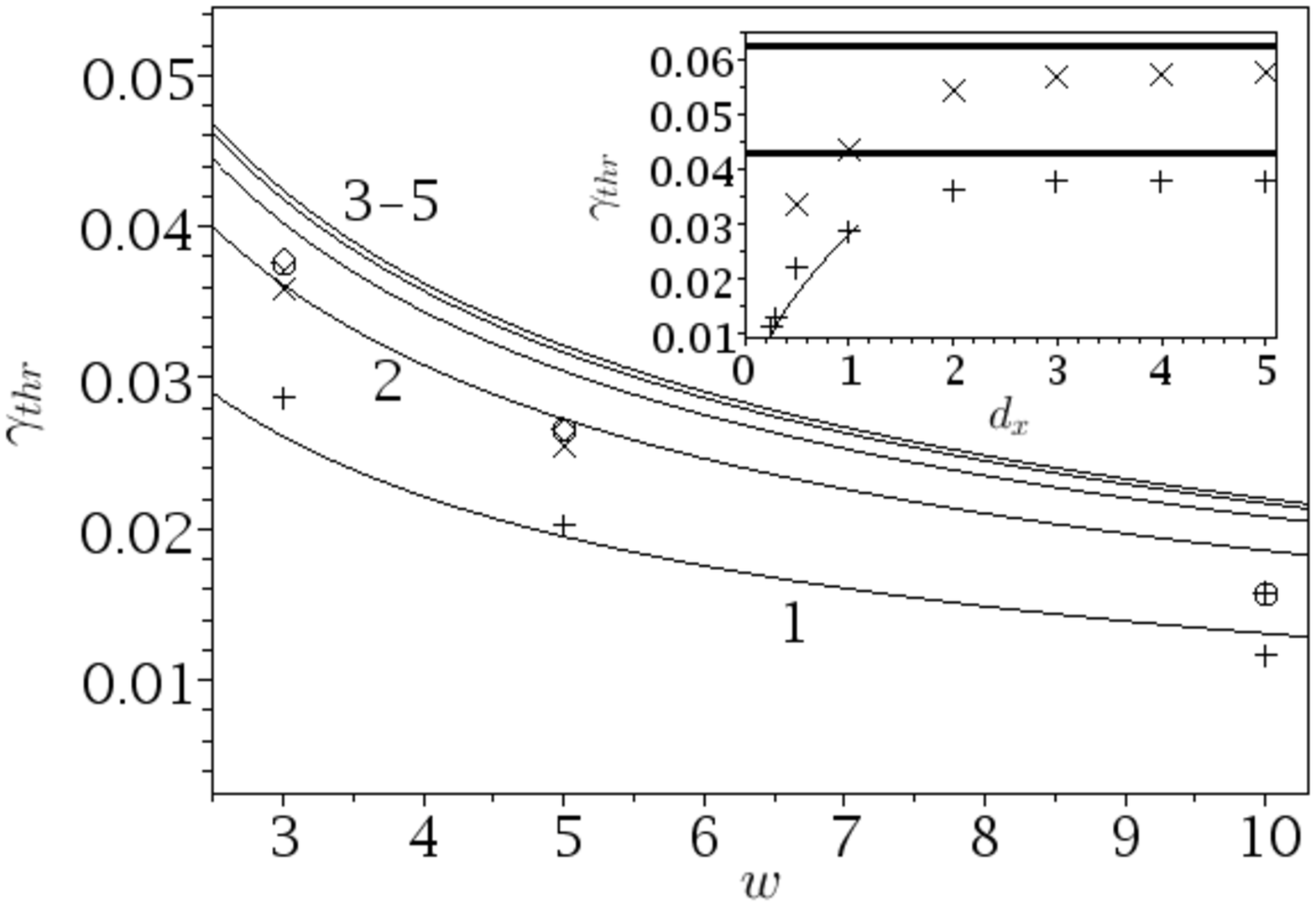,width=4.8in,angle=0}}
\caption{Retrapping current as a function of the junction width $w$
for the rectangular impurity (\ref{4}) at $\alpha=0.1$, $\mu=0.5$ and
$d_y=1$. The markers correspond to the cases $d_x=1$ ($+$), $d_x=2$ 
($\times$), $d_x=3$ ($\oplus$), $d_x=4$ ($\diamond$) and $d_x=5$ 
($\blacklozenge$).
The analytical approximation (\ref{9}) is given by the solid lines
$1-5$ for $d_x=1,2,\ldots, 5$, respectively.
The inset shows the retrapping current dependence on $d_x$
at $d_y=1$ ($+$), $d_y=2$ ($\times$) and $w=3$. Two thick solid lines
correspond to the analytical approximation (\ref{9}) at $d_x \to \infty$ 
for $d_y=1$ and $d_y=2$, thin solid line corresponds to the approximation
 (\ref{10}) at $d_y=1$.
}
\label{fig4}
\end{figure}

The results of numerical simulations given in Fig. \ref{fig4} 
demonstrate satisfactory correspondence with the analytical
approximations (\ref{9})-(\ref{10}).
%
The deviations from the numerical results appear to be stronger
as compared to the point and line impurities, mostly because
we were unable to provide the ${\cal O}(\alpha^2)$ correction that
works equally well in both limits $d_x \to 0$ and $d_x \to \infty$ and also
due to higher degree of the fluxon deformation during the interation
with the impurity. 
It should be mentioned that even for $d_x \gtrsim 3$ one can
consider the ``long'' $d_x$ limit and it works fairly well. Indeeed, 
the markers that correspond to $d_x=3$ ($\oplus$), $d_x=4$ ($\diamond$) and 
$d_x=5$ ($\blacklozenge$) seem to be almost indistinguishable in Fig. \ref{fig4}.

\section{Discussion and conclusion}

To summarize, we have shown that quasi-one-dimensional fluxon
 passage across microshorts is significantly enhanced in comparison
with the purely one-dimensional case. The retrapping threshold
current decays with the junction width approximately 
as $w^{-1/2}$, according to
the kinematic approach. The numerical simulations support this
dependence for intermediate (several $\lambda_J$) values of $w$.
With the increase of $w$, the discrepancy between the numerical results
and the kinematic approximation becomes more distinct. 
The reason of this discrepancy lies in the fact that for
$w \gg 1$ the fluxon cannot be longer considered as an completely 
rigid object and its deformation the $y$ direction should be taken
into account.

One can formulate the following simple argument that explains
the enhanced fluxon transmission across the obstacle of the
 width $d_y$ in the Q1D case. The obstacle can be described as a localized (both in the $x$ and $y$ directions)
potential barrier. Only the central part ($|y| \gtrsim d_y/2$) of the initially homogeneous  in the $y$ direction fluxon 
 takes part in the interaction process, while the marginal areas 
 $ d_y/2 \lesssim |y|\le w/2 $ do not. Thus, if the energy in the tails
 is sufficient enough to overcome the barrier, the fluxon will pass.
If $w \to \infty$, the energy in the non-interacting part of the fluxon
tends to infinity, and, consequently, it will overcome any localized
obstacle.

Finally, we would like to outline the future research directions. 
In our opinion they lie beyond the rigid rod approximation used
in this Letter. In this case the fluxon center of mass depends
also on the transverse coordinate $y$ and the effective equations
of motion are PDE's and not ODE's, like Eq.~(\ref{7}). Accounting
for the dependence on $y$ becomes important in the following
cases: (i) the junction width is too large and the flexural
oscillations along the fluxon line appear; (ii) the dissipation
is rather small and/or the junction length is $L\sim w$, thus
 the spatial distortions of the fluxon line do not have enough time
 to die out. These cases are also interesting in connection with 
 the arc-like fluxons found in the infinitely wide junction 
 \cite{m91pd} and the variety of the excitations that travel
along the fluxon crest in the dissipationless 2D junctions  \cite{gksyn}.
The combined effect of the dissipation 
and the finite junction width makes the transverse-invariant
 fluxon front the energetically most favourable
solution. However, other attractors of the system which are non-transverse-invariant may exist as suggested by Figs. \ref{1}j-l, especially
when dissipation is small. 
Finding them and investigating how they will manifest themselves on
the junction current-voltage dependence is an important problem.

\section*{Acknowledgemet}
Authors are indebted to DFFD (project F35/544-2011) for
financial support.

\end{document}